\documentclass[runningheads]{llncs}
\usepackage{makeidx}  
\usepackage{color}
\usepackage{subfig}
\usepackage{amsmath}
\usepackage{amssymb}
\usepackage{booktabs}
\usepackage{pdfsync}
\usepackage{graphicx}
\usepackage{bm}
\usepackage{anysize}
\newcommand{\mbf}[1]{{\ensuremath{\mathbf{#1}}}}

\newcommand{\comment}[1]{}

\newcommand{\normal}[2]{\mathcal{N}(#1 \given #2)}

\newcommand{\given}{\,|\,}

\newcommand{\bbeta}{\mbox{\boldmath $\beta$}}

\newcommand{\bpsi}{\mbox{\boldmath $\psi$}}

\newcommand{\boldzero}{{\bf 0}}

\newcommand{\bfx}{{\mbox{\boldmath $x$}}}
\newcommand{\bfy}{{\mbox{\boldmath $y$}}}

\newcommand{\bfC}{{\mbox{\boldmath $C$}}}
\newcommand{\bfD}{{\mbox{\boldmath $D$}}}

\newcommand{\bfI}{{\mbox{\boldmath $I$}}}

\newcommand{\bfK}{{\mbox{\boldmath $K$}}}

\newcommand{\bfV}{{\mbox{\boldmath $V$}}}

\newcommand{\bfY}{{\mbox{\boldmath $Y$}}}

\newcommand{\B}[1]{{\mbox{\boldmath $#1$}}}

\newcommand{\T}{{\top}}
\newcommand{\diag}{{\rm diag}}

\newcommand{\cut}[1]{}

\marginsize{2.6cm}{2.6cm}{2.6cm}{2.6cm}

\begin{document}
\graphicspath{{figures/}}
\newcommand{\OLI}[1]{{\color{blue} OLI: #1}}
\newcommand{\KB}[1]{{\color{blue} KARSTEN: #1}}
\newcommand{\AH}[1]{{\color{red} ANTTI: #1}}
\newcommand{\chr}[1]{{\color{magenta} CHRISTOPH: #1}}
\newcommand{\BARBARA}[1]{{\color{magenta} BARBARA: #1}}
\newcommand{\Kpop}{\bfK^{\text{pop}}}
\newcommand{\Knoise}{\bfK^{\text{noise}}}
\newcommand{\Vpop}{\bfV^{\text{pop}}}

\mainmatter              
\title{A mixed model approach for joint genetic analysis of
  alternatively spliced transcript isoforms using RNA-Seq data}
\titlerunning{Mixed model mapping for transcript eQTL}  
%
\author{Barbara Rakitsch\inst{1} \and Christoph Lippert\inst{2} \and Hande Topa\inst{3} \and
  Karsten Borgwardt\inst{1,4} \and Antti Honkela\inst{5} \and Oliver Stegle\inst{1}}
\authorrunning{Rakitsch et al.} 
\institute{Max Planck Institutes T\"ubingen, T\"ubingen, Germany,\\
\email{\{rakitsch,borgwardt,stegle\}@tuebingen.mpg.de}
\and
Microsoft Research, Los Angeles,
California,
USA \\
\email{lippert@microsoft.com}
\and
Helsinki Institute for Information Technology HIIT,
Department of Information and Computer Science,\\
Aalto University, Helsinki, Finland, \\
\email{hande.topa@aalto.fi}
\and
Zentrum fuer Bioinformatik,
Eberhard Karls Universit\"at T\"ubingen,
T\"ubingen,
Germany
\and
Helsinki Institute for Information Technology HIIT,
Department of Computer Science,\\
University of Helsinki, Helsinki, Finland, \\
\email{antti.honkela@hiit.fi}
}

\maketitle              

\begin{abstract}
RNA-Seq technology allows for studying the transcriptional state of
the cell at an unprecedented level of detail.
Beyond quantification of whole-gene expression, it is now possible to
disentangle the abundance of individual alternatively spliced transcript isoforms of a gene.
A central question is to understand the regulatory processes that
lead to differences in relative abundance variation due to external
and genetic factors.
Here, we present a mixed model approach that allows for $(i)$ \emph{joint}
analysis and genetic mapping of multiple transcript isoforms and $(ii)$ mapping of \emph{isoform-specific} effects.
Central to our approach is to comprehensively model the causes of
variation and correlation between transcript isoforms, including the
genomic background and technical quantification uncertainty.
As a result, our method allows to accurately test for shared as well as
transcript-specific genetic regulation of transcript isoforms and
achieves substantially improved calibration of these statistical
tests.
Experiments on genotype and RNA-Seq data from 126 human HapMap
individuals demonstrate that our model can help to obtain a more
fine-grained picture of the genetic basis of gene expression variation.
\end{abstract}

\newpage
\section{Introduction}
\label{sec:introduction}
{\bf Motivation} Large-scale genotyping and expression profiling initiatives
have fostered expression quantitative trait loci (eQTL)
analyses, investigating the genetic component of the transcriptional
state of the cell.
In human, eQTL studies have revealed an abundance of genetic associations
between proximal polymorphic loci and individual
genes~\cite{Stranger2007,stranger2012patterns,pickrell2010understanding,montgomery2010transcriptome}.  
In model organisms, such as segregating yeast strains~\cite{Brem2002,Smith2008},
mouse~\cite{schadt_integrative_2005} or
{\it{Arabidopsis thaliana}}~\cite{west2007global},
studies have uncovered a map of the genetic component of gene
regulation, which is characterized by an interplay of local genetic
associations in {\it cis} and distal genetic effects with {\it trans}
mechanisms.
The majority of existing studies of transcription were based on
microarray technologies, which are limited to a coarse quantification
of the total transcript abundance of {\it a-priori} known genes
present in a sample.

Only recently, thanks to second generation sequencing techniques, deep
transcriptome sequencing (RNA-Seq) has become viable even for
population-scale analyses.
Seminal work~\cite{montgomery2010transcriptome,pickrell2010understanding} has
demonstrated the merits of using the digital readout provided by
RNA-Seq technologies for genetic analyses, allowing for a more
comprehensive dissection of the genetic component of transcriptional
variability.
RNA sequencing not only allows for an unbiased genome-wide
quantification of transcription at an unprecedented resolution, it
also enables the detection and quantification of alternative splice
forms of genes~\cite{mortazavi_rna-seq_2008,wang_RNA-Seq_2009}.
Differences in the abundances of individual isoforms of a gene have
been shown to play an important role for phenotypes in specific
tissues.
For example, genetic modifications that affect alternative splicing and
post-transcriptional quality control mechanisms are postulated to be
involved in a large proportion of genetic disorders, including the
development of cancer~\cite{matlin2005understanding,skotheim2007alternative,he2009global,fackenthal2008aberrant,danckwardt2002abnormally}.

By treating transcript isoform abundances as quantitative traits it is
possible to perform eQTL studies on the level of individual isoforms.
Such an approach allows for detecting polymorphic loci affecting the
expression level of specific isoforms, for example via regulation of
splice factors.
While there have been previous attempts to tackle this problem at the level of
individual
exons~\cite{montgomery2010transcriptome,pickrell2010understanding},
there is a lack of statistical models that allow for accurately
dissecting the genetics of transcriptional gene-regulation
at the level of individual transcript isoforms.
Such methods face various challenges that need to be addressed in
order to get meaningful results: First, the expression levels of
individual transcript isoforms are highly correlated due to common
transcriptional regulation as well as shared exons and splicing
mechanisms between different isoforms.
Second, limited identifiability of individual transcript isoforms from
RNA-seq data causes correlation between isoform abundance estimates, which
can confound the analysis if not accounted for.
Finally, as in almost any genomic analyses of quantitative traits,
there is the need to account for structure within the sample, for
example due to shared ancestry or population clustering.

{\it Our goal in this article is to present a probabilistic model that
  allows for detecting (i) genetic loci that affect the
  overall expression level of multiple transcript isoforms and (ii) loci that act
  in an isoform-specific manner, while addressing the aforementioned challenges.}

{\bf Related approaches} An alternative paradigm to what we present
here are approaches that perform the analysis on the level of single
exons instead of complete isoforms.
Such analyses approaches have been used in the context of genetic
perturbations~\cite{montgomery2010transcriptome,pickrell2010understanding}
and other factors~\cite{anders2012detecting}.
While such an approach imposes fewer assumptions on the processes that
are involved, individual exon-specific regulation needs to be
retrospectively combined to a transcript-level view.
Furthermore, sensitive transcript expression quantification can borrow
statistical strength across exons of the same transcript, potentially
picking up more subtle variation (See also discussions
in~\cite{Glaus2012,katz2010analysis}).


 Conceptually, our approach is related to multi-trait mixed
 models, which have undergone a long history of development (see
 e.g.~\cite{henderson1976multiple,stich2008multi,price2011single}).
 Recently, these approaches have been successfully applied to
 genome-wide association studies of correlated traits and have been
 shown to increase power to detect pleiotropic SNP effects on two
 correlated traits as well as being able to detect
 environment-specific effects of SNPs on a phenotype measured under
 two different environments~\cite{korte2012mixed}.

{\bf Contributions of this paper}
In this work we propose a mixed model to jointly map
individual isoforms from alternatively spliced genes and to uncover
genomic variants that affect \emph{all} or \emph{specific} transcript isoforms.
 By employing a state-of-the-art approach to infer expression levels
 of individual isoforms~\cite{Glaus2012}, we build a problem-specific
 background covariance that explains confounding genetic and technical
 correlation between transcript isoform abundance estimates of the
 same gene.
 We apply the resulting mixed model to search for proximal \emph{cis}
 associations in 126 HapMap individual where we find extensive
 regulation at gene level but also genetic effects that alter the
 expression abundance in a transcript-specific manner.

 Besides a fine-grained picture of the genetic basis of gene
 expression, another key advantage of the proposed approach is that it
 is convenient to accurately measure statistical significance in our
 model.
 $p$-values for association can be computed analytically through
 likelihood ratio tests, without the need for expensive permutation
 experiments to assess statistical significance.
 By controlling for sources of confounding our model does not suffer
 from $p$-value inflation, as it is observed in simpler methods, and
 thus provides better control for the type-1 error rate.


%
%
\section{Transcript expression inference from RNA-Seq}
\label{sec:transcr-expr-infer}
Transcript isoform expression can be only indirectly reconstructed
from RNA-Seq datasets.
Here, we build on BitSeq~\cite{Glaus2012}, a fully probabilistic
approach to estimate transcript isoform expression levels from RNA-Seq
data and a known transcript annotation.

Briefly, the generative model of the RNA-seq reads in each sample
assumes a categorial assignment of reads to one of all genome-wide
transcripts or a noise category according to
$p(I_{j}^n \given \mbf{\theta}_{\cdot}^n) =
\text{Cat}(I_{j}^n \given \{\theta_{1}^n,\dots,\theta_{M}^n\})$,
where $\theta_{t}^n$ is the probability of observing a fragment of
transcript isoform $t$ in sample $n$, and $M$ denotes the
total number of annotated transcript isoforms.
The likelihood of a read $r_j$ given a transcript assignment,
$p(r_j \given I_j = m) = p(r_s \given seq_{mps})p(p \given m)p(s
\given m)$, is given by
the probability of drawing a read from the specific position in
the transcript isoform $m$ while accounting for mismatches, position and
sequence bias correction.
Given an observed set of reads, we can apply
Bayesian Markov chain Monte Carlo inference to
obtain a full posterior probability distribution over the
transcript isoform relative expression levels $\theta_{t}^n$.
These are transformed to log-expression levels $y_{t}^n = \log \theta_{t}^n$
for further analysis.

For each gene, we use the Monte Carlo estimates to obtain summary
statistics that capture the mean expression level and covariation
between the abundance estimates $y_{t}^n$ of the $T_g$ different
transcripts of each gene $g$ within each sample $n$.  Modeling
covariation is useful because transcript isoforms often share sequence
which causes ambiguity in assigning reads to individual isoforms, and the
transcript expression posteriors often show strong negative
correlation (See also~\cite{Glaus2012}).
These transcript isoform means and
between-isoform expression covariance summaries
are then used for further modeling.
This first and second moment of the variation for each gene is
explained by a Gaussian over $\bfy^n = \{y^n_{1},\dots,y^n_{T_g}\}, p(\bfy
\given \mathcal{D}) =
\normal{\bfy}{\hat{\bfy}, \bfC^{n}}$.

Along the same lines, the same Monte Carlo estimates can be use to
estimate whole-gene expression, ignoring the transcript structure.
For a specific gene $g$ in sample $n$, expression levels can be
estimated as $y_g^n = \log  \sum_{t\in T} \theta^n_t$,
where $T$ is the set of all transcripts belonging to the gene
$g$.
These estimate yield RKPM estimates that are near-identical to
standard counting-based methods.

\section{A multi-isoform mixed model}
\label{sec:multi-transcr-mixed}
To differentiate joint effects from isoform-specific regulation it
is important to account for different sources of correlation between
multiple transcript isoforms of the same gene.
Here, we propose a multi-isoform variance component model that
comprehensively accounts for different origins of correlation.
First, multiple transcript isoforms of the same gene are correlated
across samples because of genetic factors that affect multiple
transcripts in the same way.
In the absence of transcript-specific splicing regulation, this
variation is dominated by a common gene expression component.
Second, there is technical covariation between the isoforms quantified
in the same sample (see Section~\ref{sec:transcr-expr-infer}).

In Section~\ref{sec:multi-trait-variance}, we introduce a
linear mixed model that accounts for both of these types of genetic and
technical factors that induce correlation between transcripts.
In Section~\ref{sec:testing-sqtl-effects}, we show how the model can be used to test for common and specific genetic
effects of a SNP (single nucleotide polymorphism).

\subsection{Multi-isoform linear mixed model}
\label{sec:multi-trait-variance}
For a specific gene (we drop the dependence on $g$ to unclutter notation), let the $N\cdot T$-dimensional vector $\bfY =
[\bfy_1,\dots,\bfy_T]$ denote the vector of log expression levels for $T$
different isoforms of a single gene measured in each of the $N$ samples.
For each $t\in[1,\dots,T]$, we model the $N$-dimensional vector
$\bfy_t=(y_{t}^{1},\dots,y_{t}^{N})$ holding all samples of isoform $t$ by a linear model as
follows:
\begin{align}
  \bfy_t = \underbrace{\B{1}\cdot\mu_t}_{\text{Transcript mean}} +\underbrace{\bfx^{\star} \cdot\beta^{\star}}_{\text{{\it SNP} effect}}+ \underbrace{\sum_{l \in
    \text{pop}} \bfx_l \cdot\beta_{l,t}}_{\text{population structure}}
  +  \underbrace{\bpsi_t}_{\text{noise}}.
\end{align}
Here, $\mu_t$ is the bias for transcript $t$, $\bfx^{\star}$ is a SNP
we would like to test for association, for example in the vicinity of
the gene.
The SNPs  $\{\bfx_l\}_{l \in  \text{pop}}$ are genome-wide markers
that capture population structure, excluding SNPs within a window of
$\pm1$ mega base around the gene to avoid proximal
contamination~\cite{listgarten2012improved} when testing SNPs that lie
in {\it cis}.
The genetic \textit{cis} effect $\beta^{\star}$ is modeled as a fixed
effect.
The population structure effects $\{\beta_{l,t} \}_{l \in \text{pop}}$
are modeled as random.
To allow for correlation between the effects of population structure
across all isoforms, the $T$-dimensional vector, $\bbeta_l =
[\beta_{l,1},\dots,\beta_{l,T}]$  is coupled across transcript
isoforms by an identical multivariate normal distribution with covariance
$\Vpop$, for each SNP $\bfx_l$ used to represent population structure.
\begin{align}
  \bbeta_l \sim \mathcal{N}(\boldzero, \Vpop);
  \;\;\;\;\;\;\;\;
  \Vpop =
  \left[
  \begin{array}{ccc}
  v^{\text{pop}}_{1,1}&\;\dots\;&v^{\text{pop}}_{1,T}
  \\
  \vdots & \ddots &\vdots
  \\
  v^{\text{pop}}_{1,T} & \dots &v^{\text{pop}}_{T,T}
  \end{array}
  \right]
\end{align}
The $T(T+1)/2$ independent entries in $\Vpop$ are inferred from the
data within the model.
For genes with a large number of transcript isoforms, the number of parameters
may become prohibitive, in which shrinkage approaches such as Lasso
regulation on the inverse covariance~\cite{friedman2008sparse} could be used; see also Discussion.

The observational noise is independent across individuals, but
correlates across isoforms within each individual due to
technical variation.
For every individual $n$, we let the $T$-dimensional vector $\bpsi^n =
[\psi_{1}^n,\dots,\psi_{T}^n]$ be the noise across all isoforms, which
is distributed as
\begin{eqnarray}
\bpsi^n \sim \mathcal{N}(\boldzero,\diag\left({\B\delta}\right) + \alpha^2 \bfC^n),
\end{eqnarray}
where $\B{\delta} = [\delta_1^2,\dots,\delta_T^2]$ is residual noise level of each isoform,
$\bfC^n$ denotes the technical correlation matrix between the isoforms
due to quantification in sample $n$ and $\alpha^2$ is a common scaling
parameter of the technical covariation.
The technical correlation matrices $\bfC^n$ are estimated {\it a-priori} by
the BitSeq algorithm (see Section~\ref{sec:transcr-expr-infer}).
The noise-variances $[\delta_1^2,\dots,\delta_T^2]$, and $\alpha^2$
are estimated within the model.

\begin{figure}[h!]
    \begin{center}
    \includegraphics[width=0.8\textwidth]{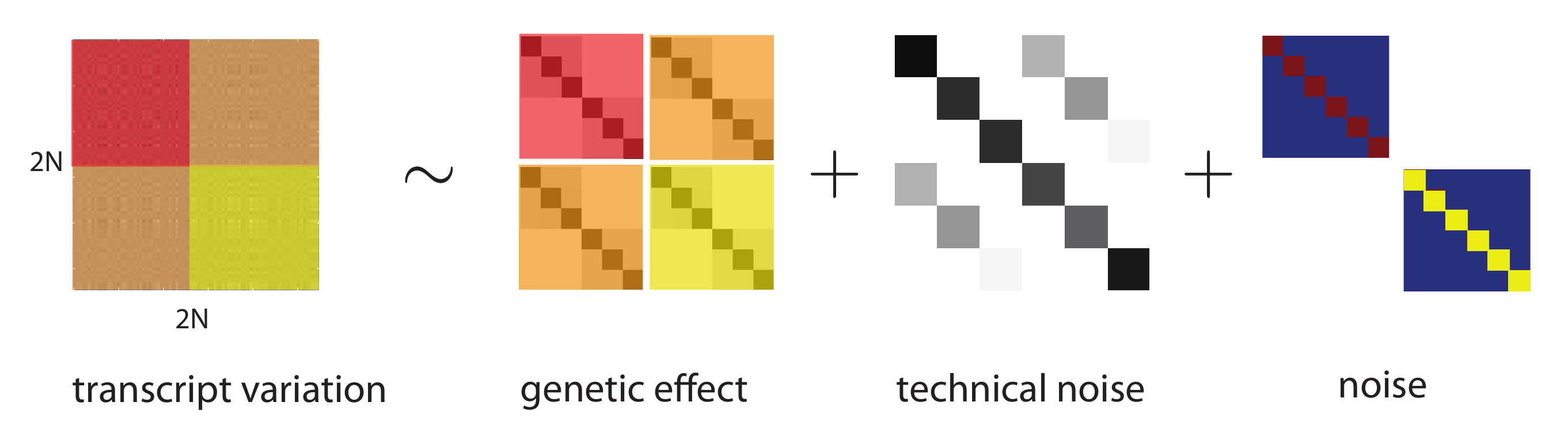}
    \end{center}
  \caption{Illustration of the marginal covariance model for a gene
    with two transcript isoforms. The total covariance of the stacked
    transcript expression profiles are decomposed into a genetic
    component, technical covariance and transcript-specific
    observation noise.
  }
  \label{fig:covariance_illustration}
\end{figure}

\paragraph{Marginal model}
Integrating over the effects of population structure $\beta_{l,t}$ for all $l$ and
$t$ we get a joint marginal likelihood for all isoform levels of the gene:
\begin{align}
  \left[
    \begin{array}{c}
      \bfy_1\\
      \vdots\\
      \bfy_T
    \end{array}
  \right]\sim
  \mathcal{N}\left(
    \left[
      \begin{array}{c}
        \B{1}\cdot\mu_1\\
        \vdots\\
        \B{1}\cdot\mu_T
      \end{array}
    \right]\;;\;
    \Kpop
    +
    \Knoise
  \right),
  \label{eq:multi_marginal_lik}
\end{align}
where the covariance term consists of the genotype signal covariances
\begin{align}\nonumber
  \Kpop=
  \left[
    \begin{array}{ccc}
      v_{1,1} \left(\sum_{l \in
          \text{pop}} \bfx_l\bfx_l^{\T}\right) & \;\hdots\; &v_{1,T}\left(\sum_{l \in
          \text{pop}} \bfx_l\bfx_l^{\T}\right)
      \\
      \vdots & \ddots&\vdots
      \\
      v_{1,T}\left(\sum_{l \in
          \text{pop}} \bfx_l\bfx_l^{\T}\right)& \;\hdots\; & v_{T,T}\left(\sum_{l \in
          \text{pop}} \bfx_l\bfx_l^{\T}\right)
    \end{array}
  \right],
\end{align}
and the noise covariance
\begin{align}\nonumber
\Knoise =
  \alpha^2 \underbrace{\left[
    \begin{array}{ccc}
      \bfD_{1,1}  & \;\hdots\; & \bfD_{1,T}
      \\
      \vdots & \ddots&\vdots
      \\
      \bfD_{T,1}& \;\hdots\; & \bfD_{T,T}
    \end{array}
  \right]}_{\text{technical noise}}
  +
   \underbrace{\left[
    \begin{array}{ccc}
      \delta_{1}^2 \bfI  &  &\boldzero
      \\
      & \ddots&
      \\
      \boldzero&  & \delta_{T}^2 \bfI
    \end{array}
  \right]}_{\text{noise}},
\end{align}
where $\bfD_{i,j} = \diag([C^1_{i,j}, \dots, C^N_{i,j}])$.
Note that the noise covariance $\Knoise$ is a sum of a diagonal matrix
for observation noise and a chessboard pattern matrix consisting of
diagonal matrices that couple different transcript within individual
samples $n$ due to technical noise.
For an illustration of the effective covariance for a simple
two-transcript gene, see Figure~\ref{fig:covariance_illustration}.
Note that in the case of a gene with a single transcript, this
variance component collapses to a standard linear mixed model with
realized relationship matrix (see for example~\cite{Kang2010,fastlmm}).

We estimate the variance parameters of the model by maximizing the
marginal likelihood (\ref{eq:multi_marginal_lik}) using constrained
quasi-Newton optimization. To avoid prohibitive computations due to
re-estimation of covariance parameters for different {\it cis} SNPs
$\bfx^{\star}$, we estimate all parameters only once for each gene,
ignoring the fixed effect of the {\it cis} SNP,
Analogous approximations, fitting the background covariance parameters
on the null model, have previously been successfully applied for
GWAS~\cite{Kang2010}.

\subsection{Mixed model to test for common and transcript isoform-specific
  genetic effects}
\label{sec:testing-sqtl-effects}
We would like to test for the effect of individual {\it cis} SNPs that
either have a common effect on all transcript isoforms or act in an
isoform-specific manner.
For this purpose, we use the parameters obtained from the
multi-isoform variance components model Eqn.~\eqref{eq:multi_marginal_lik} in
uni-variate testing.
We construct a linear mixed model using different designs modeling the effect of $\bfx^{\star}$ as either common genetic effects across transcript isoforms or
isoform-specific genetic effects:

\begin{align}
\left[
\begin{array}{c}
\bfy_1-\B{1}\cdot\mu_1\\
\vdots\\
\bfy_t-\B{1}\cdot\mu_t\\
\vdots\\
\bfy_T-\B{1}\cdot\mu_T
\end{array}
\right]
\sim
\mathcal{N}\left(
\left[
\begin{array}{c}
\B{1}\\
\vdots\\
\B{1}\\
\vdots\\
\B{1}
\end{array}
\right]\cdot b
+
\underbrace{
\left[
\begin{array}{c}
\bfx^{\star}\\
\vdots\\
\bfx^{\star}\\
\vdots\\
\bfx^{\star}
\end{array}
\right]
\cdot\beta_0}_{\text{joint effect}}
+
\underbrace{
\left[
\begin{array}{c}
\boldzero\\
\vdots\\
\bfx^{\star}\\
\vdots\\
\boldzero
\end{array}
\right]
\cdot\beta^{\star}}_{\text{specific effect}}
\;;\;
\sigma_c^{2}\underbrace{\left(\Kpop
+
\Knoise \right)}_{\text{relatedness matrix}}
+\sigma_e^2\bfI_{N \cdot T}
 \right).
\label{eq:mixed_model}
\end{align}
Here, the joint effect component is the identical SNP $\bfx^{\star}$ replicated over each transcript isoform.
From this linear mixed model, different likelihood ratio tests can be
carried out for each SNP in {\it cis}, using the computational tricks
proposed in the FaST-LMM algorithm~\cite{fastlmm}.
In the implementation, we specify the relatedness matrix to be the sum
of $\Kpop$ and $\Knoise$ estimated on the null model.
This approach enables us to obtain $p$-values from a $\chi^2$
distribution.
Related testing strategies, however focusing on pairs of quantitative
traits in GWAS, have been previously used in~\cite{korte2012mixed}.

\paragraph{Gene-level association test}
In order to test for \emph{joint effects} of SNPs that act on transcription on a gene-level, namely $\bfx^{\star}$ that jointly regulate all transcript isoforms in the same direction, we fit the mixed model in Equation~\eqref{eq:mixed_model} only with the joint effect, discarding transcript-specific effects.
This alternative model is compared to a null model by additionally dropping the joint effect and performing a one degree of freedom likelihood ratio test.

\paragraph{Isoform-specific association test}
In order to test for \emph{specific effects} of SNPs, i.e. whether a SNP $\bfx^{\star}$ has an effect that acts specifically on a transcript isoform $t$, we fit the full mixed model in Equation~\eqref{eq:mixed_model}, placing the specific effect only on the abundance of this isoform $\bfy_t$.
 This alternative model is compared to a null model that drops the specific effect. By conditioning on the joint effects we ensure that we only retrieve effects that act differently between the transcript isoforms. This results in a likelihood ratio test with one degree of freedom.

\paragraph{Combined association test}
Finally, we consider a multi-transcript test that assesses
whether there is \textit{any} association with a transcript isoform $t$, either by jointly regulating all isoforms or by acting specifically on the isoform $t$.
The alternative model is the same as when testing for isoform-specific tests. However, in this case we drop both, the joint effect as well as the specific effect from the null model.
In this case the likelihood ratio test has two degrees of freedom.

In the following we will denote these three testing strategies
when using the multi transcript covariance structure fitted on the
null model (Section~\ref{sec:multi-trait-variance}) as multi-isoform
mixed model (MIMM).
We also consider alternative methods for the purpose of comparison,
for example without the multi trait structure and the technical noise
covariance.

\section{\emph{cis} QTL mapping of transcript isoform levels in
  HapMap populations}
\label{sec:experiments}
We applied the multi-isoform mixed model (MIMM) and alternative methods toeinstellbar
previously published RNA-Seq data of human HapMap
samples~\cite{altshuler2010integrating}, combining the data from
Pickrell et. al~\cite{pickrell2010understanding} and Montgomery
et. al~\cite{montgomery2010transcriptome}.
Altogether, there were  59 samples from the CEU population and
74 samples from the YRI HapMap population that could be mapped to
unique HapMap identifiers.
SNPs were filtered for minimum allele frequency 0.05 and we discarded variants with
missing genotyping information in more than 5\% of the samples.
Finally, we also discarded all samples with more than 5\% missing SNPs.
After filtering, our dataset comprised of a total of 126 individuals
and 996,158 SNPs.

Transcript quantification was done using BitSeq
(Section~\ref{sec:transcr-expr-infer}), where we used the GRCh37.p68
reference transcriptome from Ensembl and mapped the RNA-Seq
reads with bowtie~\cite{Langmead2009}.
As the focus of this work resides on transcript isoform-specific effects, we
only considered genes with at least two transcript isoforms.
We used the marginal uncertainty estimates from BitSeq
(Section~\ref{sec:transcr-expr-infer}) to filter out transcript isoforms that
were consistently not expressed or difficult to quantify in 90\% of
the samples or more ($z$-score cutoff 1.5).
Because of the small sample size in these datasets, genes with more
than 4 active isoforms were discarded, ensuring that the comparison
of methods was not compromised by sample size.
This final dataset consisted of~5,954 genes, comprising of a total
number of 16,340 individual transcripts.

Statistical tests for proximal {\it cis} effects were carried out in a
1 Mb window upstream or downstream of each gene start and stop.
In both the MIMM and ordinary mixed models, we considered for
comparison, we cut out this region for building the multi-isoform
relationship matrix $\Kpop$ to avoid possible loss of power due to
proximal contamination~\cite{listgarten2012improved}.
We also included an indicator variable distinguishing between the YRI
and CEU population to account for this low-rank component of
population structure.
Estimates of a local false discovery rate for each gene were obtained
from the Benjamini \& Hochberg~\cite{benjamini1995controlling}
procedure to estimate $q$-values.

\subsection{Testing for general regulatory effects on whole gene
  expression and transcript isoform levels}
\label{sec:test-gener-regul}
\begin{figure}[ht!]
    \begin{center}
    \subfloat[][LMM gene-level Effects (red) vs. MIMM combined
    association test (blue)]{
      \includegraphics[width=0.25\textwidth]{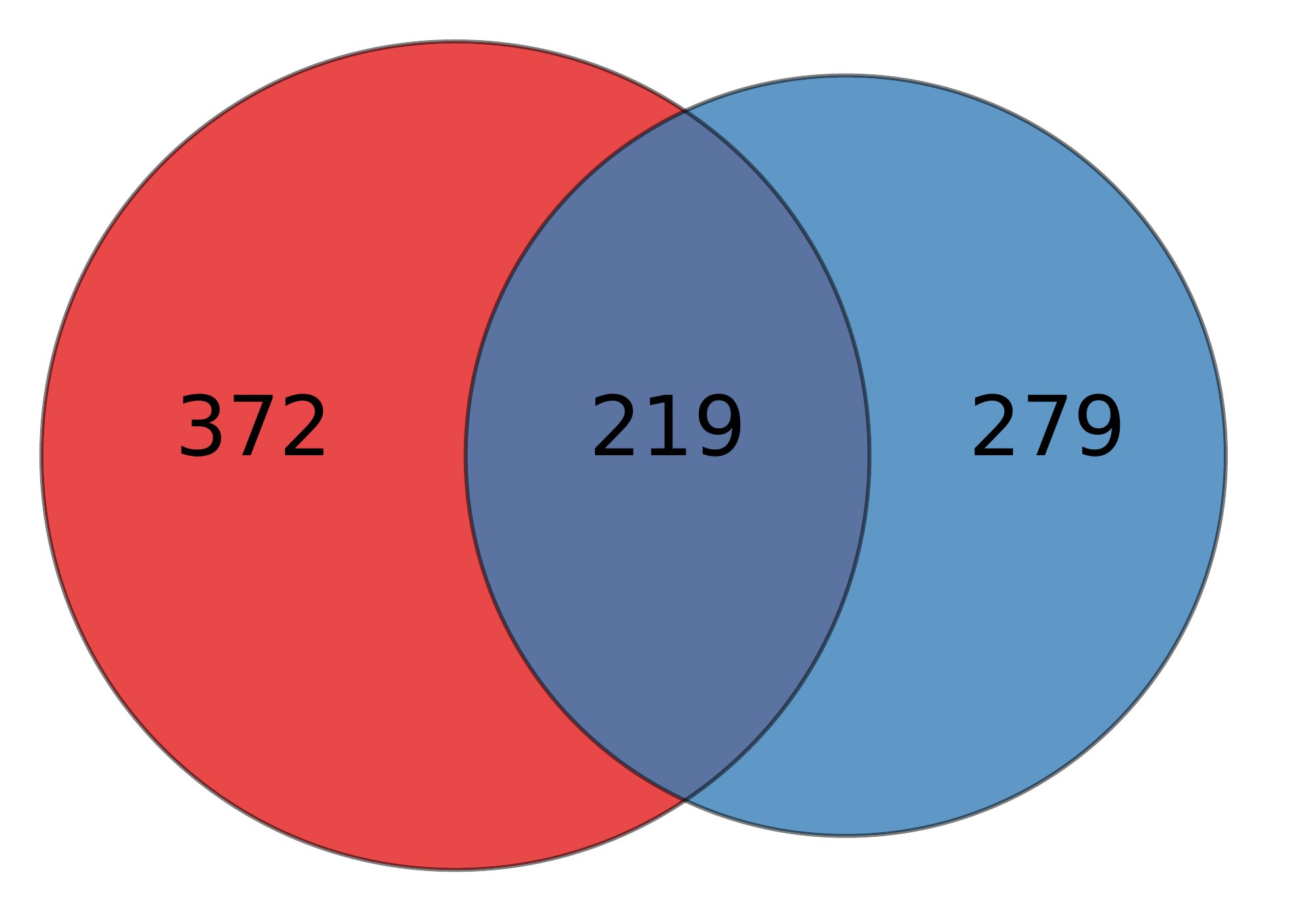}
    }
    \subfloat[][Linear Mixed Model: whole gene expression]{
      \includegraphics[width=0.25\textwidth]{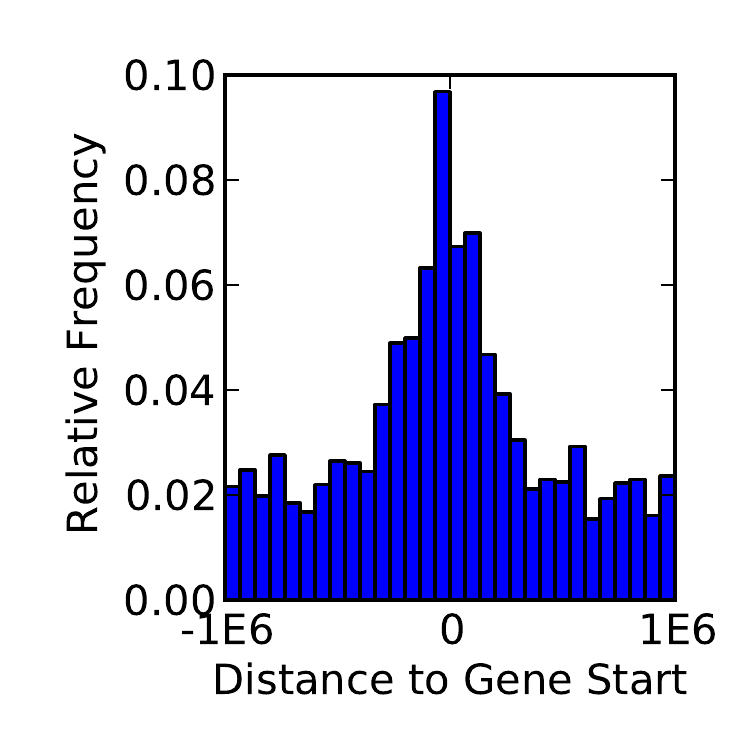}
    }
    \subfloat[][Multi-isoform Mixed Model: combined association test]{
      \includegraphics[width=0.25\textwidth]{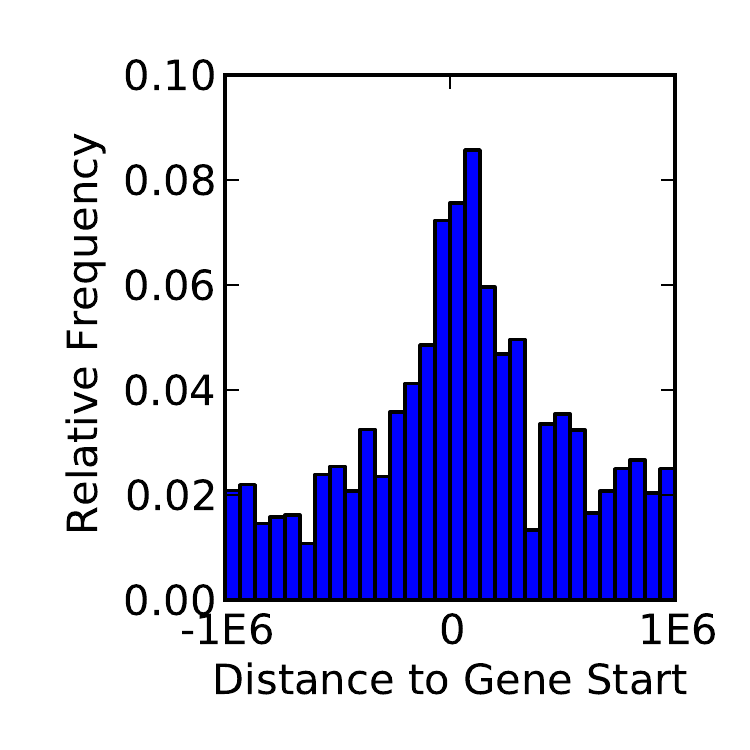}
    }
    \end{center}
    \caption{Comparison of significant results ($q$-value threshold 5\%) of the MIMM
      model, testing for combined associations, and a standard linear
      mixed model applied to whole gene expression levels.
      \textbf{(a)} Mutual overlap of associations.
      \textbf{(b-c)} Density of the distance to the gene for
      \emph{cis} associations.
    }
  \label{fig:MMT_versus_gene}
\end{figure}

First, we wanted to relate results obtained by transcript-based
modeling, as proposed here, with associations from an eQTL scan on whole
gene expression, as considered in previous
analyses~\cite{montgomery2010transcriptome,pickrell2010understanding}.
To this end, we applied the multi-isoform mixed model (MIMM) and
performed the combined association test to find genetic effects that
are either specific to individual transcript isoforms or act on the
gene-level.
We compared these results with a linear mixed model (LMM), carrying
out standard eQTL tests on whole gene expression levels that were also
estimates from BitSeq in a manner consistent with the transcript level
estimates (Section~\ref{sec:transcr-expr-infer}).
Many but not all genes that had at least one significant \emph{cis}
association by one of the methods ($q$-value threshold of 5$\%$) were also
detected by the other approach (see Figure~\ref{fig:MMT_versus_gene}
(a)).
This result suggests that the combined association test on a transcript
level is capable of retrieving associations that correspond to
established eQTLs but also other signals.
As expected, both models yielded associations with a strong enrichment
for lying in close proximity to the gene start
(Figures~\ref{fig:MMT_versus_gene} (b) and (c)), suggesting that many
of these hits are genuine QTLs.
We also carried out genome-wide scans for a random selection of 49 of
genes to investigate calibration of $p$-values, finding that both methods
yielded calibrated $p$-values (genomic control: $\lambda=1.02$ MIMM, $\lambda=1.03$ LMM).
Summary results are shown in Table~\ref{tab:association_summary}).

\begin{table}[htbp]
   \centering
   \resizebox{\linewidth}{!}{
    \begin{tabular}{l|rrr|rrr|rrr|rrr} 
        &\multicolumn{3}{c|}{\# significant genes} &
    \multicolumn{3}{c|}{\# significant isoforms} &
        \multicolumn{3}{c|}{\# significant SNPs} &
    \multicolumn{3}{c}{Genomic Control ($\lambda$)}\\
\hline\hline
& \bf \;\;LIN&\bf \;\;LMM&\bf MIMM & \bf \;\;LIN&\bf \;\;LMM&\bf MIMM &\bf \;\;LIN\bf &\bf \;\;LMM&\bf MIMM
&\bf \;\;LIN\bf &\bf \;\;LMM&\bf MIMM \\\hline
\multicolumn{13}{c}{\it Gene expression tests} \\

    general genetic effect& 612 & 591 & -&-&-& - & 6417 &  6011 & -
    &1.03 & 1.03 & -\\\hline
    \multicolumn{13}{c}{\it Transcript expression tests} \\
combined genetic effect &  4,085 & 3,608 &  498 & 11,302 & 9,000& 1,016 &
  992,017 &   975,583 & 10,521 &
  1.37 & 1.51 & 1.02
   \\
{isoform-specific effects}
& 2,497 & 3,498 &  429 & 6,110  & 8,430 &  707&
    991,041 &   1,226,504 & 4,315   &
    1.12 & 1.73 & 1.00
    \\
{gene-level effects} & 2,581 & 453 &  441  & &&&
 88,494 & 18,546 & 4,516  &
 1.42 & 1.02 & 1.02
 \\
\multicolumn{7}{c}{}
 \end{tabular}
 }
    \caption{Summary statistics of the associations retrieved by
      alternative methods on different expression traits.
      Top row: Gene expression test, carried out on whole-gene
      expression estimates ignoring transcript structure.
      Bottom row: Transcript expression tests, carried out on all
      transcripts within a gene jointly using alternative testing
      strategies. }
   \label{tab:association_summary}
\end{table}

\subsection{Dissecting genetic associations specific to individual
  transcript isoforms}
\begin{figure}[ht!]
\centering
\includegraphics[width=0.4\textwidth]{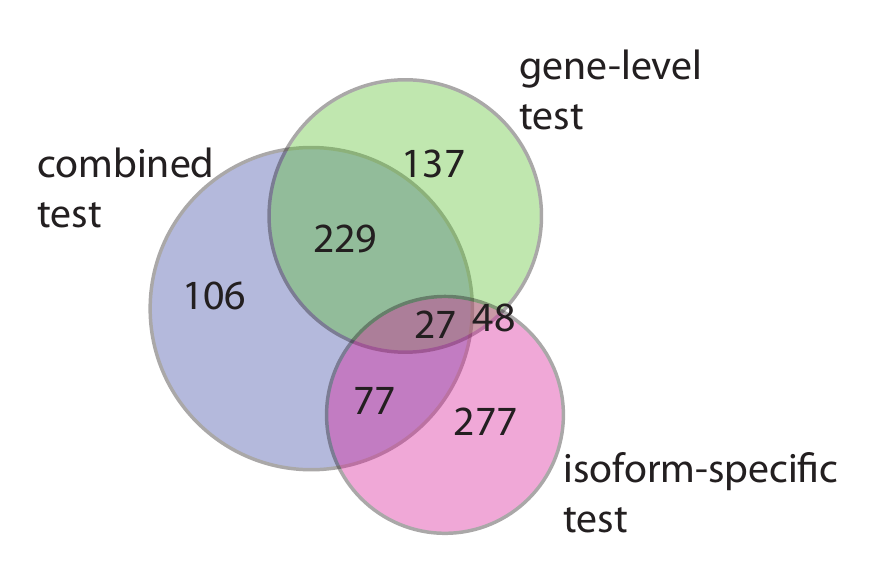}
\caption{Overlap of associations retrieved by the multi-isoform mixed
  model when considering alternative testing strategies.
}
\label{fig:MIMM_overlap}
\end{figure}

Next, we strived to dissect the set of associations retrieved by the
combined association test into associations that are either common to
all transcripts, i.e. occur at a gene-level, and transcript-specific
genetic effects.
To this end, we used the MIMM and additionally carried out
gene-level association tests and isoform-specific association tests in
the same local \emph{cis} regions as before.
The overlap of results obtained from the three alternative testing
strategies are shown in Figure~\ref{fig:MIMM_overlap}.
The combined association test retrieved the greatest number of genes
with at least one association, overlapping with many of the hits from
either the gene-level test or the transcript level test.
Notably, genes with an associations found by the gene-level
test and the transcript-specific test were almost completely disjoint,
suggesting that within individual genes there is a strong tendency for
either gene-level regulation or transcript-level regulation.

\begin{figure}[h!]%
\begin{centering}
  \subfloat[][QQ-plot  Linear Model]{
\includegraphics[width=0.3\textwidth]{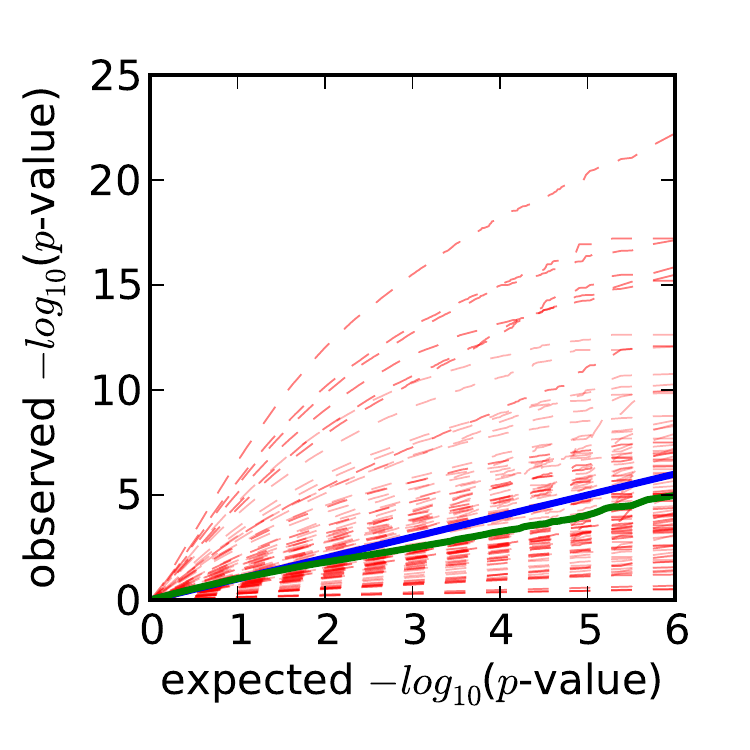}
  }
  \subfloat[][QQ-plot Linear Mixed Model]{
\includegraphics[width=0.3\textwidth]{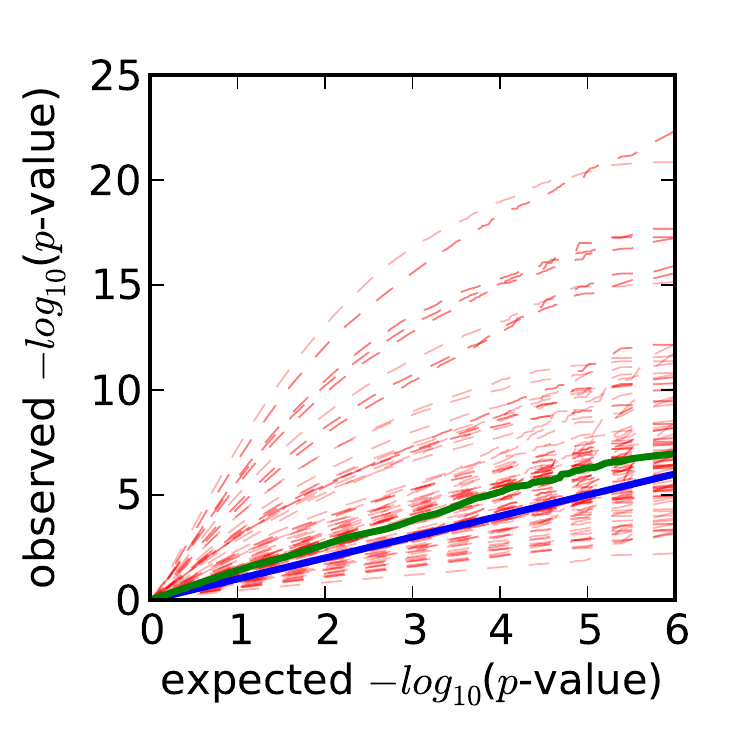}
  }
   \subfloat[][QQ-plot Multi-isoform Mixed Model]{
 \includegraphics[width=0.3\textwidth]{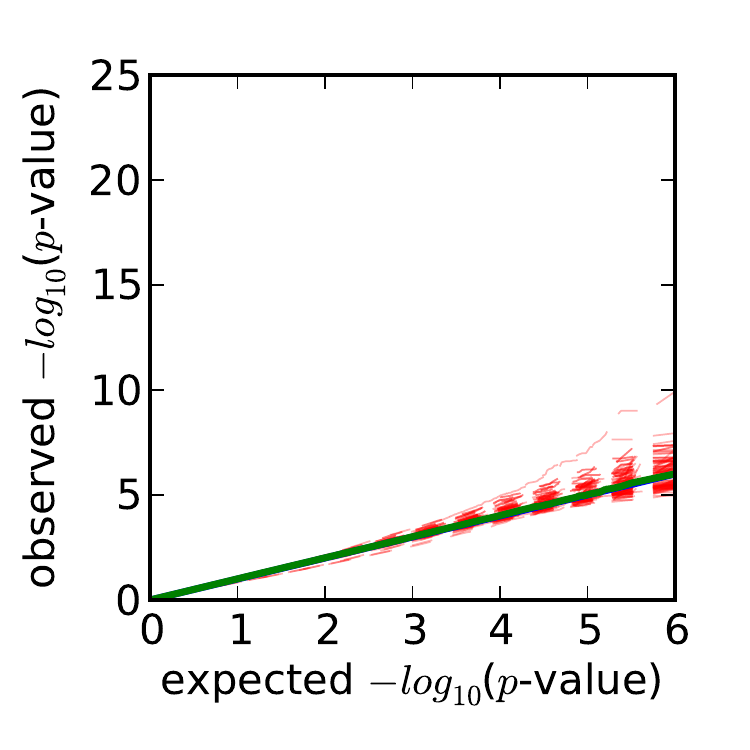}
  }
\\
  \subfloat[][Association distance: Linear Model]{
\includegraphics[width=0.3\textwidth]{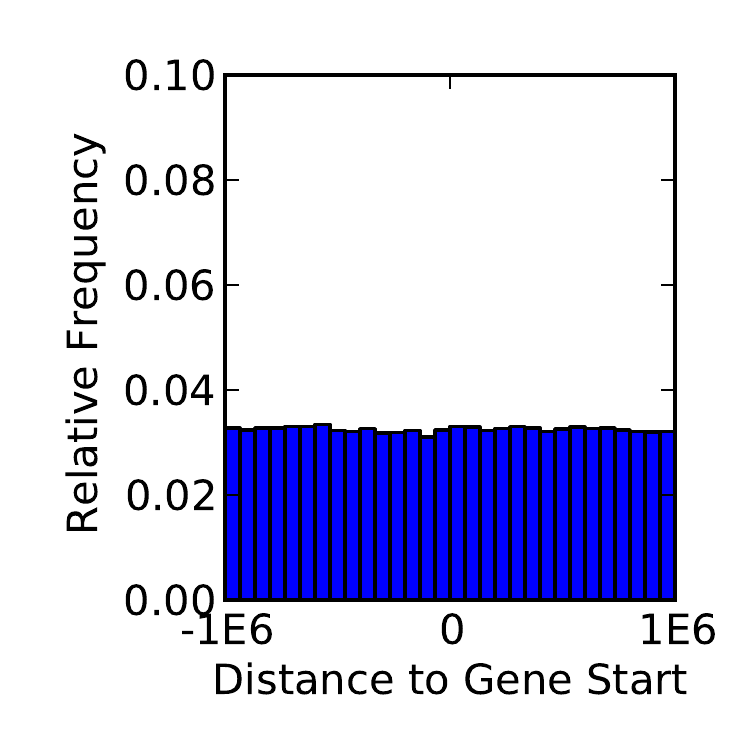}
  }
  \subfloat[][Association distance: Linear Mixed Model]{
\includegraphics[width=0.3\textwidth]{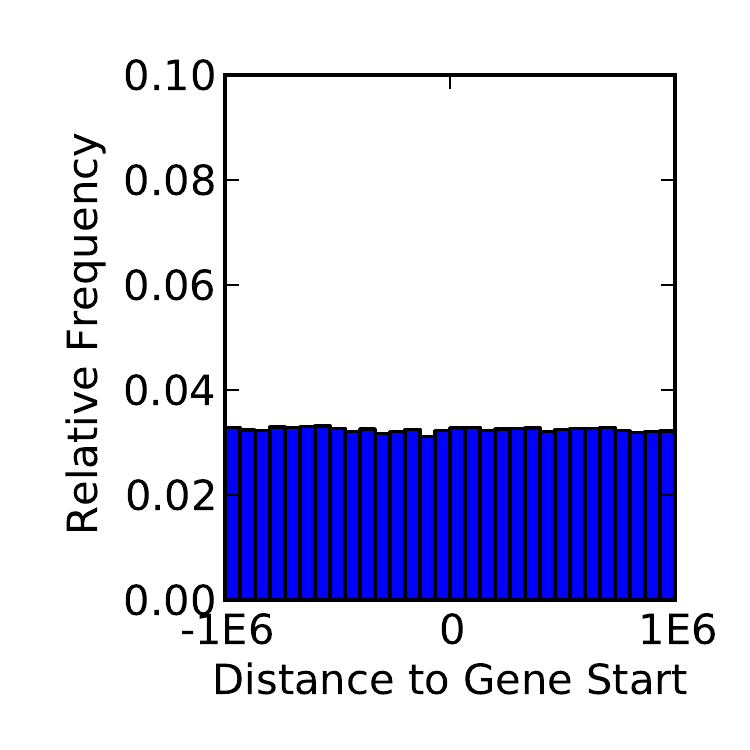}
  }
   \subfloat[][Association distance: Multi-isoform Mixed Model]{
 \includegraphics[width=0.3\textwidth]{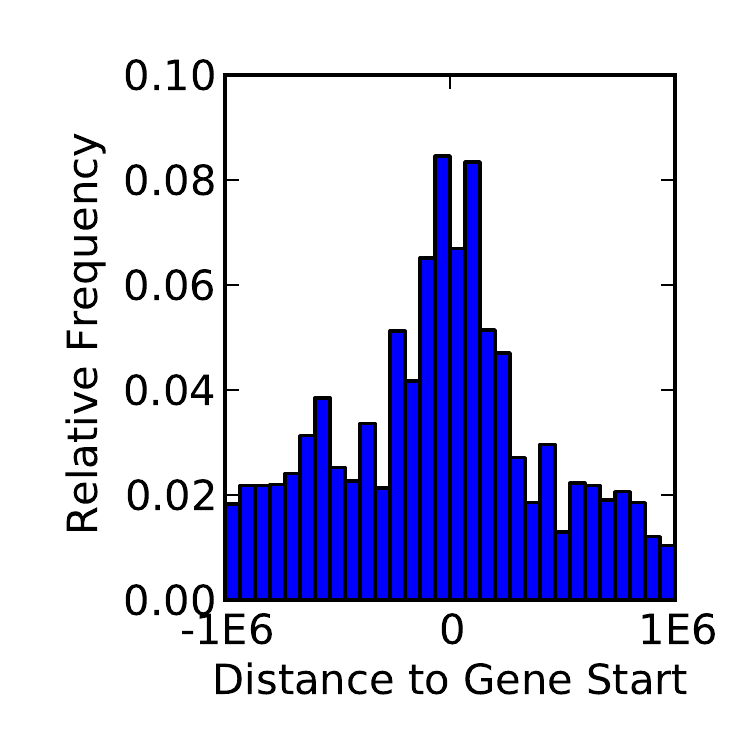}
  }
  \end{centering}
     \caption{
       Comparison of alternative methods to detect transcript-specific
       associations in proximal \emph{cis} regions.
       \textbf{(a)-(c)}: Quantile-quantile plots for genome-wide scans
       of a selection of  121 traits using alternative
       methods.
       \textbf{(d)-(f)}: Histogram of the distance of the most
       associated variant to the transcription start cite for
       significant hits (FDR 5\%).
       The multi-isoform mixed model achieved the best calibration of
       test statistics and retrieved \emph{cis} associations with an
       enrichment near the transcription start site.
     }
  \label{fig:transcript_specific}
  \vspace{-0.5cm}
\end{figure}

Perhaps the most relevant use of multi transcript-isoform models is
the association analysis of isoform-specific regulatory effects.
Figure~\ref{fig:transcript_specific} (f) shows the distance of the 707
isoform-specific associations relative to the transcript start site,
showing a similar enrichment as observed for gene-level associations
(Section~\ref{sec:test-gener-regul}).
Again, we considered a selection of transcripts and carried out
genome-wide scans to investigate calibration of p-values
(Figure~\ref{fig:transcript_specific} (c)).
This apparent calibration of tests ($\lambda = 1.00$) when using the MIMM is
crucially linked to the fitted multi trait covariance structure,
accounting for transcript correlation and different sources of noise
(Section~\ref{sec:multi-trait-variance}).
To explore the impact of this modeling, we also considered simpler
alternative covariance structure, either as used in a standard linear
mixed model (LMM) or a linear model without any background
covariance (LIN).
At a $q$-value cutoff of 5\%, the linear model and the linear
mixed model retrieved a greater number of genes with at least one
transcript-specific association (Table~\ref{tab:association_summary}).
However, these absolute numbers need to be put in context with the
statistical calibration of different methods:
Figure~\ref{fig:transcript_specific}\textbf{a-c} show genome-wide
QQ-plots for a random selection of  121 transcripts (49 genes) tested for
association by each of the considered methods.
The excess of small $p$-values retrieved by the linear model ($\lambda=1.12$) and the
linear mixed model ($\lambda=1.73$) suggests that the large number of findings is due
to severely inflated the test statistics and hence the majority of
these hits are likely to be false positives rather than actual true associations.
This hypothesis is further supported by the lack of enrichment of
associations in proximity to the transcription start site
(Figure~\ref{fig:transcript_specific}\textbf{d-f}, where only the
multi-isoform mixed model yielded hits consistent with the hypothesis
that true regulatory variants are likely to be within the gene body.
Similar deficiencies were also found when testing for joint genetic
effects (See Figure 5).

We also investigated the distribution of the number of
isoform-specific associations per gene.
The majority of genes had either a single transcripts that was under
\emph{cis} genetic regulation (38.00\%) or had two regulatory events
(59.67\%).
Only a small minority of genes had more complex regulation involving
more than two transcripts (2.33\%).

Two example genes, one with a joint genetic effect and a
transcript-specific effect are shown in
Figure~\ref{fig:example_plots}.
This illustrations demonstrates how different testing approaches
provide a finer-grained picture of the genetic regulation of gene
expression, both at the gene- and at the transcript level.

\begin{figure}[t!]%
\begin{centering}
  \subfloat[][QQ-plot  Linear Model]{
    \includegraphics[width=0.3\textwidth]{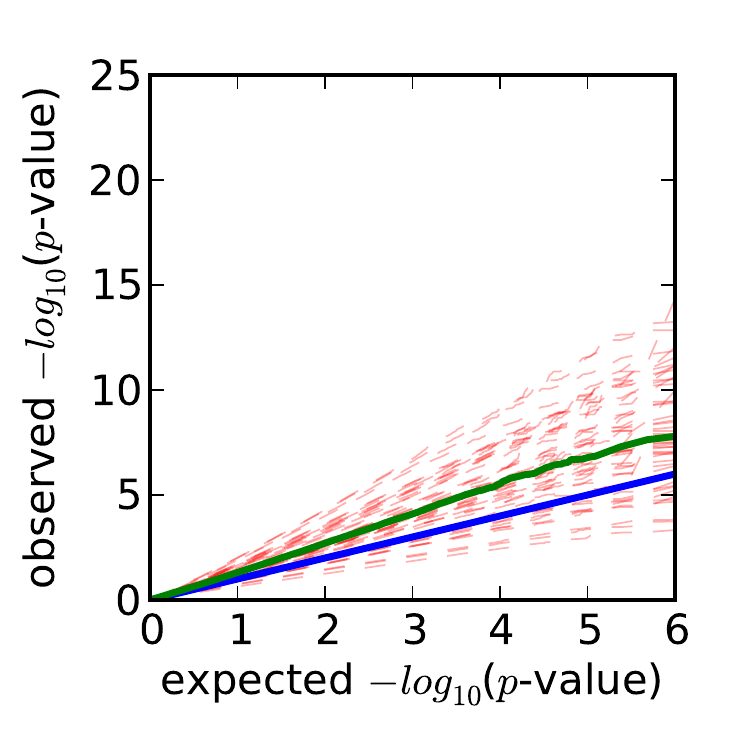}
  }
  \subfloat[][ QQ-plot Linear Mixed Model]{
    \includegraphics[width=0.3\textwidth]{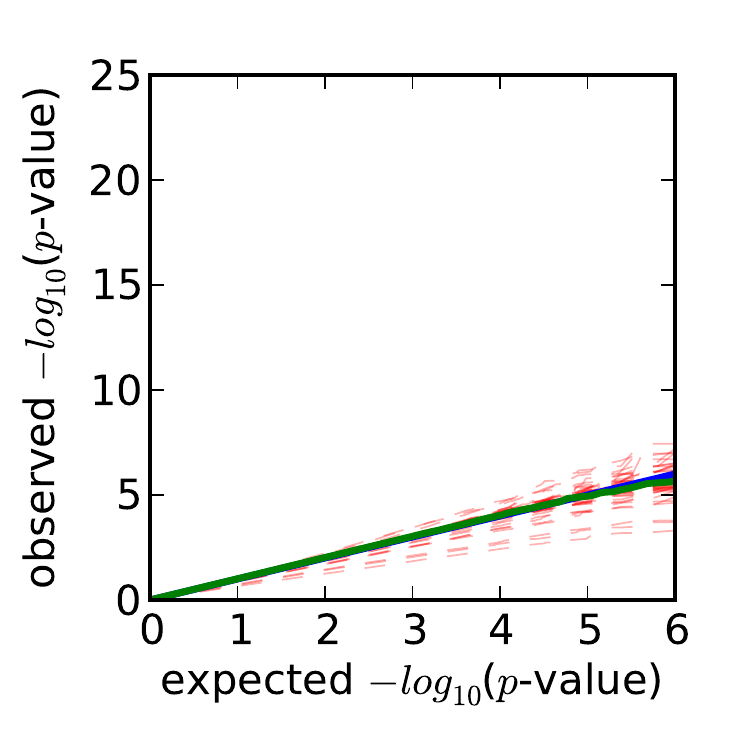}
  }
  \subfloat[][QQ-plot Multi-isoform Mixed Model]{
    \includegraphics[width=0.3\textwidth]{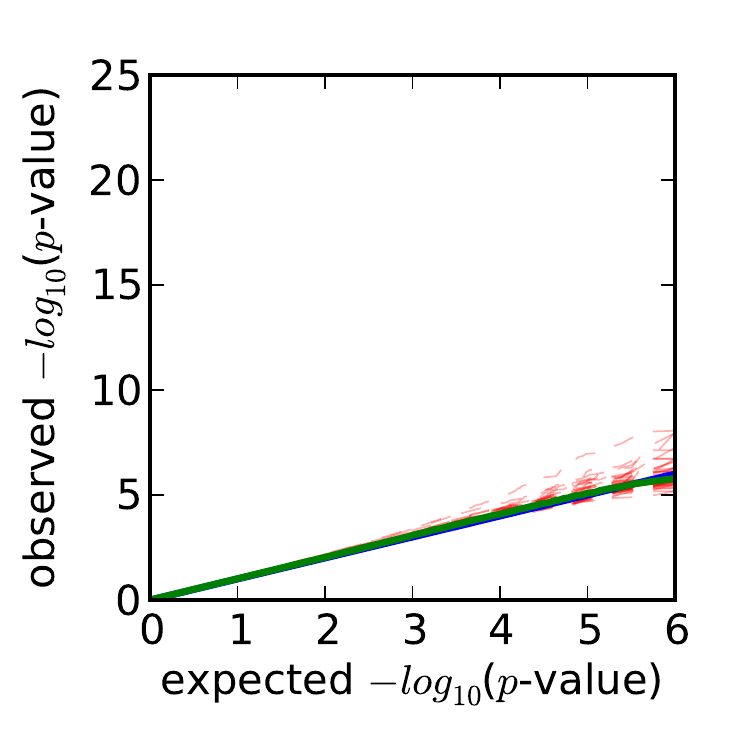}
  }
  \\
  \subfloat[][Association distance: Linear Model]{
    \includegraphics[width=0.3\textwidth]{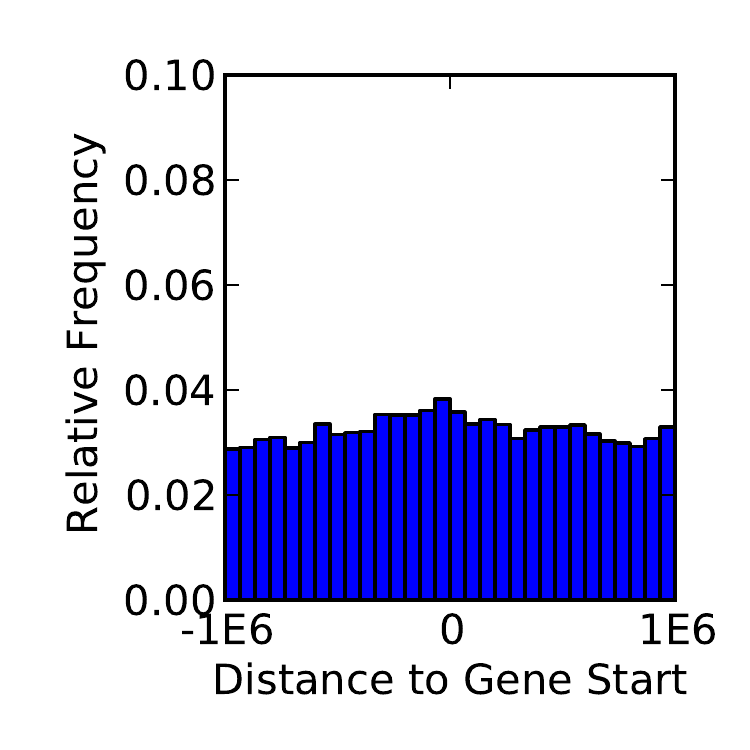}
  }
  \subfloat[][Association distance: Linear Mixed Model]{
    \includegraphics[width=0.3\textwidth]{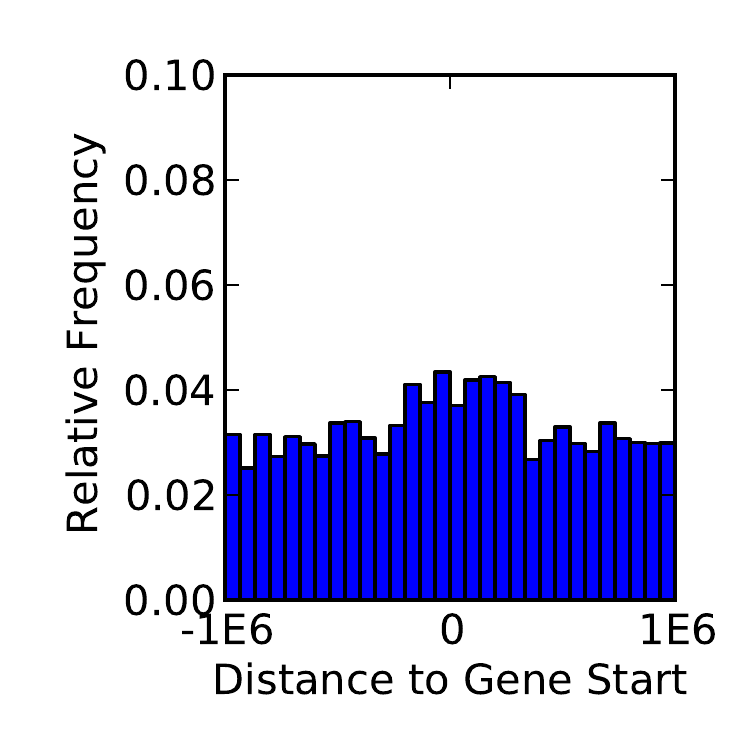}
  }
  \subfloat[][Association distance: Multi-isoform Mixed Model]{
    \includegraphics[width=0.3\textwidth]{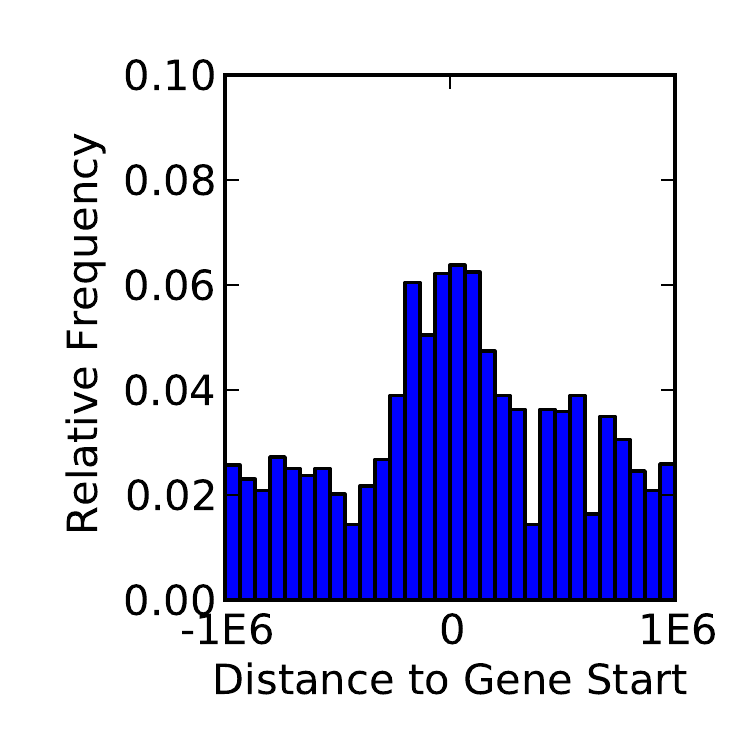}
  }
\end{centering}
\caption{
  Comparison of alternative methods to detect joint associations
  in proximal \emph{cis} regions.
  \textbf{(a)-(c)}: Quantile-quantile plots for genome-wide scans
  of a selection of  121 traits using alternative
  methods.
  \textbf{(d)-(f)}: Histogram of the distance of the most
  associated variant to the transcription start cite for
  significant hits (FDR 5\%).
  The linear mixed model and the multi-isoform mixed model showed improved calibration
  over the linear model, while the multi-isoform mixed model detected most \emph{cis} associations
  near the transcription start site.
}
  \vspace{-0.5cm}
\label{fig:gene_level}
\end{figure}

\begin{figure}[h!]

\newcommand{\geneidOne}{ENSG00000181031}
\newcommand{\geneidTwo}{ENSG00000162551}
\begin{centering}
  \subfloat[][Exons of the gene ENSG00000181031]{
   \includegraphics[width=.5\textwidth]{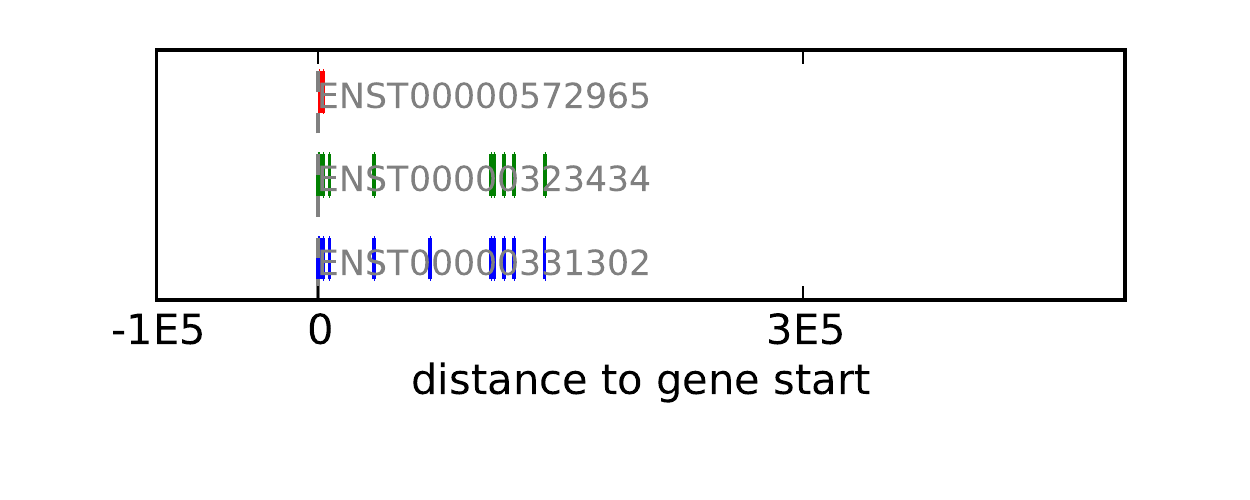}
  }
 \subfloat[][Exons of the gene ENSG00000162551]{
   \includegraphics[width=.5\textwidth]{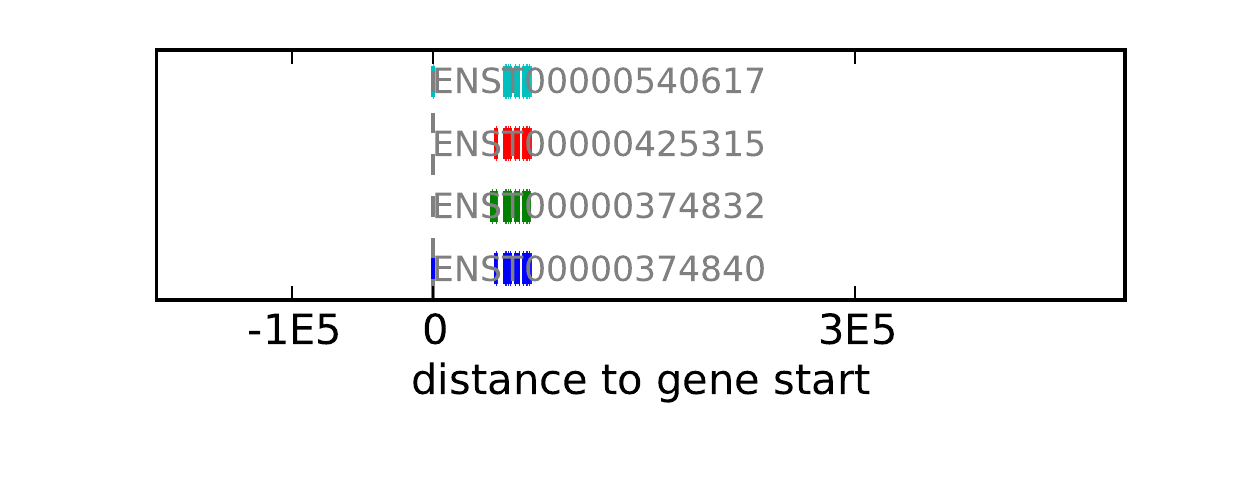}
  }
  \\
  \subfloat[][Combined association test]{
   \includegraphics[width=.5\textwidth]{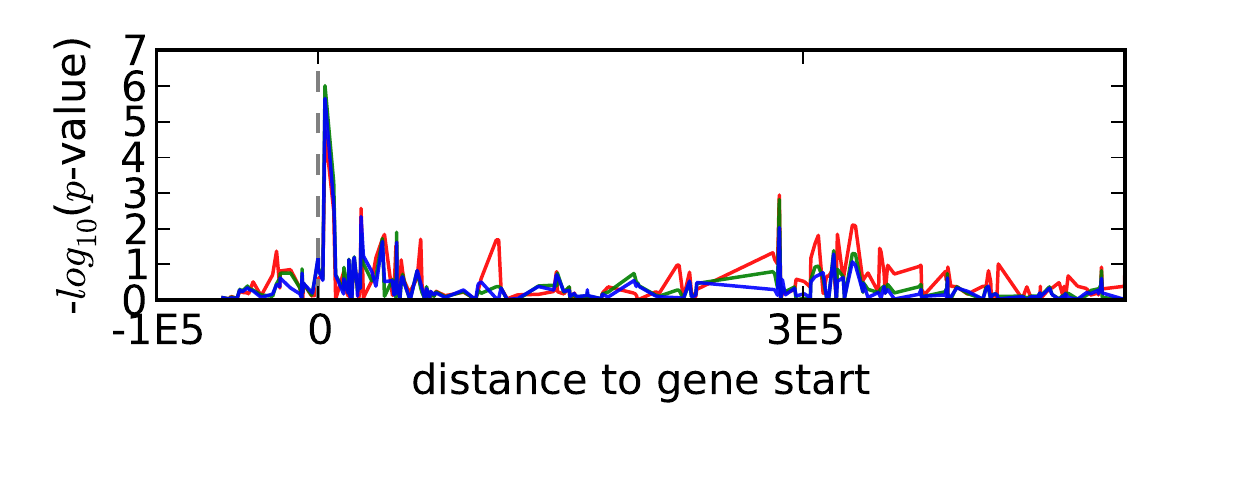}
  }
 \subfloat[][Combined association test]{
   \includegraphics[width=.5\textwidth]{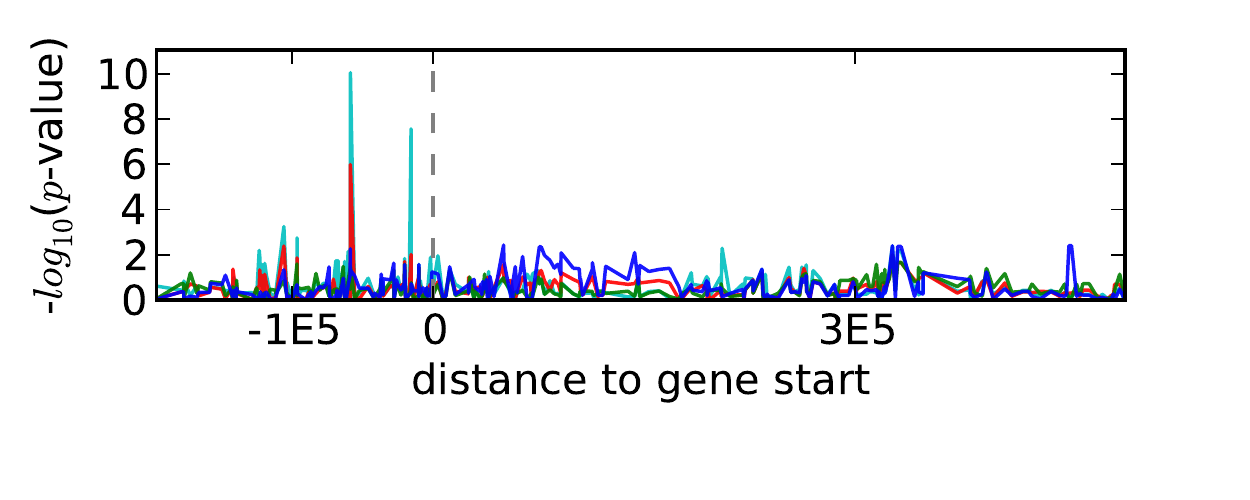}
  }
  \hfill
 \subfloat[][Joint association test]{
   \includegraphics[width=.5\textwidth]{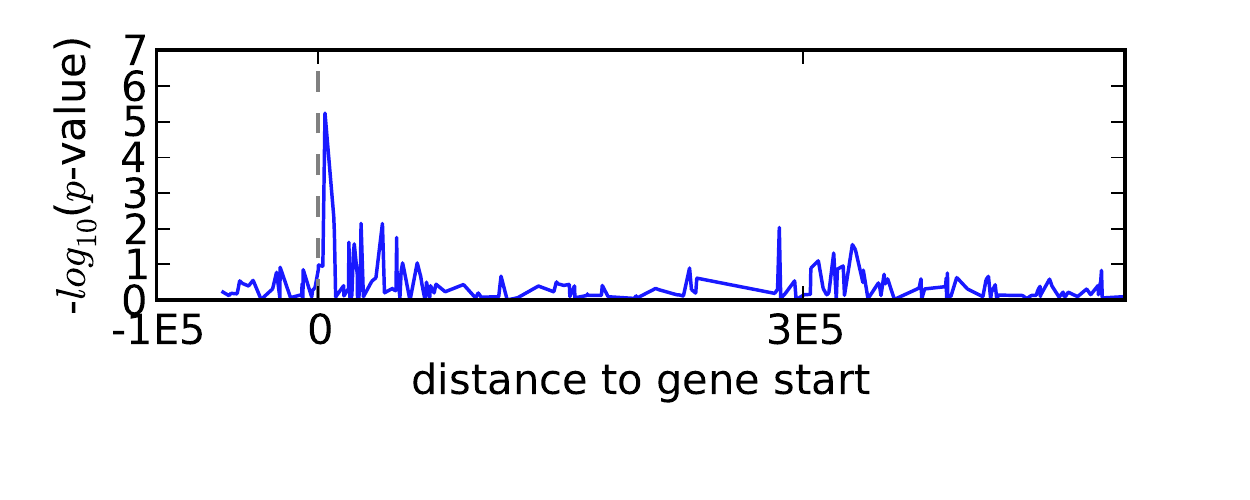}
  }
 \subfloat[][Joint association test]{

   \includegraphics[width=.5\textwidth]{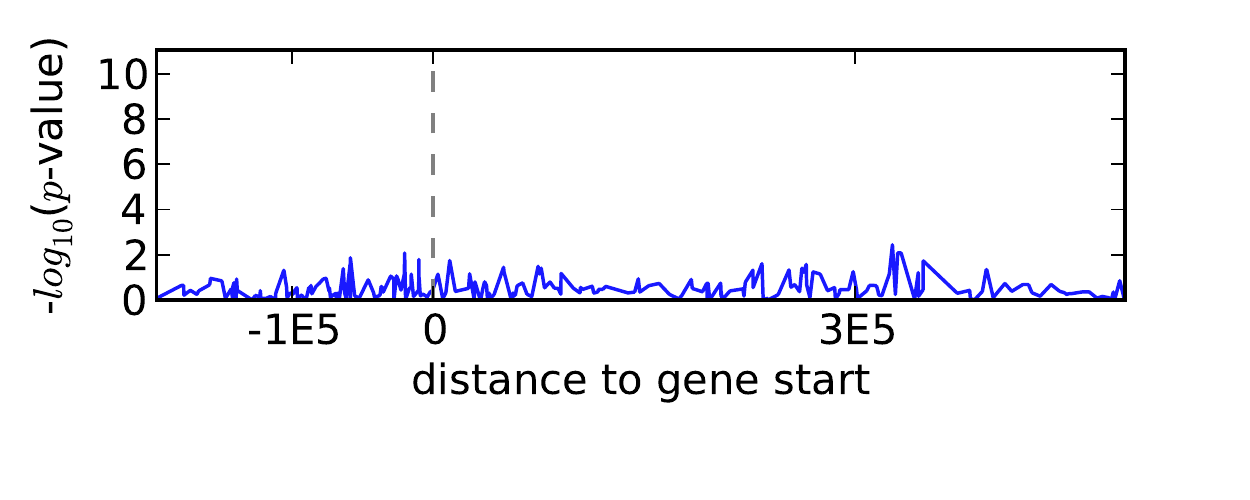}
  }
  \\
 \subfloat[][Isoform-specific association test ]{
   \includegraphics[width=.5\textwidth]{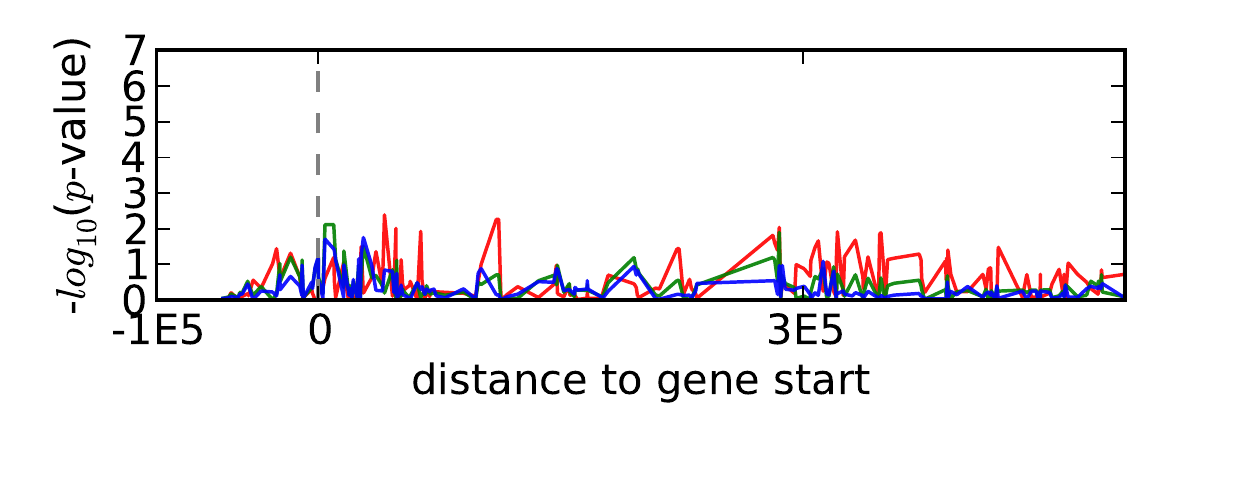}
  }
 \subfloat[][Isoform-specific association test ]{
   \includegraphics[width=.5\textwidth]{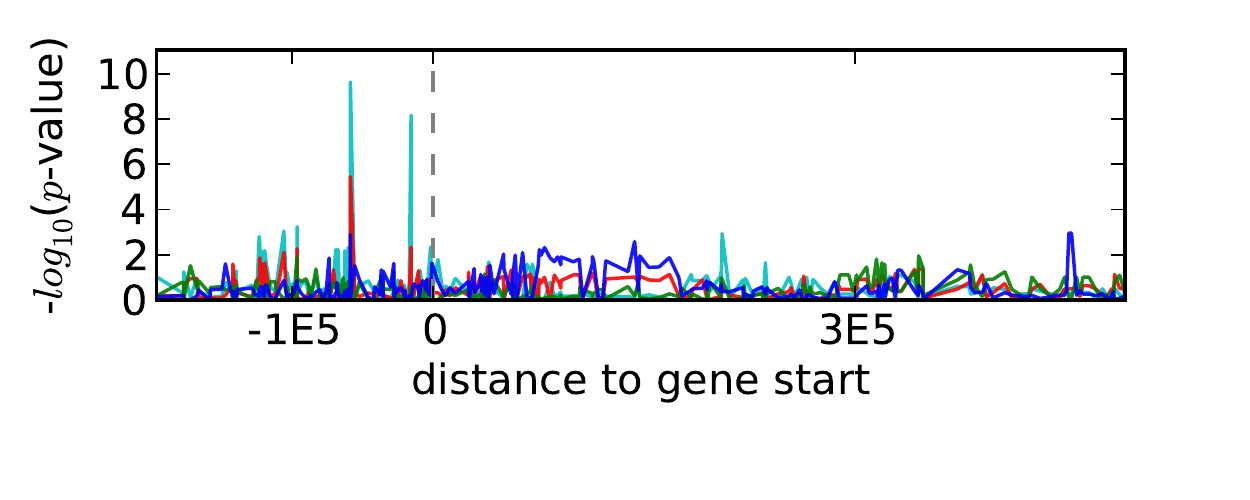}
   }
     \end{centering}
  \caption{
  Left side: Gene \textit{rph3al} (ENSG00000181031) has
   an joint association for all three transcripts
    ENST00000331302(blue), ENST00000323434(green), ENST00000572965(red).
  Right side: Gene \textit{alpl} (ENSG00000162551) has two distinct transcript-specific association for the transcript
  ENST00000540617 (cyan). The first one is shared with the transcript ENST00000425315 (red), while the second is not.  The other two transcripts show no significant association. \\
  }
\label{fig:example_plots}
\end{figure}

\section{Conclusions}

\label{sec:conclusions}
Transcriptional and post-transcriptional regulation are complex
biological processes, both of which are under genetic control.
Here, we have introduced a mixed model approach to test for genetic
effects that either act on all transcripts \emph{jointly} or alter expression
levels of \emph{specific} transcript isoforms while others remain
unaltered.

In a proof of concept application to RNA-Seq profiles from 126 HapMap
samples, we have demonstrated substantial benefits of this approach compared
to established analysis strategies.
First, we have shown that accounting for covariation between
transcripts within the same gene, both due to genetic and technical
factors, greatly improves statistical calibration of $p$-values for different
types of association tests.
The mixed model covariance structure we propose is tailored to to the
analysis of estimates of multiple transcript isoform abundances.
While general multi-trait mixed modeling has a long history for joint
modeling of multiple traits
(e.g.~\cite{henderson1976multiple,stich2008multi,price2011single,korte2012mixed}),
we are not aware of any practical application to larger sets than
pairs of traits.

Second, we have shown how different testing strategies on the level of
transcript isoforms yield more detailed mechanistic insights compared
to existing approaches that operate on a gene level, while not loosing
power.
Using statistical tests that correspond to specific regulatory
hypotheses, we were able to dissect this catalog of general \emph{cis}
associations into common genetic effects operating
\emph{across transcripts} and genetic effects that act in a
\emph{transcript-specific} manner.
While common regulatory effects are most frequent, we have found
evidence for transcript-specific regulation in 7.21\% of genes; in
some instances both types of effect were even found within the same
gene.

In conclusion, we have shown how the combination of multi-trait mixed
models and probabilistic transcript quantification is able to uncover
novel biological insights while providing statistically robust estimates.
Currently available RNA-Seq datasets are limited, both in sample sizes
and in terms of read lengths for reliable isoform quantification.
Because of these limitations, we focused on medium-complexity genes
with 2--4 transcripts.
More complex transcript structures may be fit by employing shrinkage
priors, regularizing the effective number of parameters in the
model~\cite{friedman2008sparse,StegleLMLB2012}.
In the future, larger datasets of better quality will render these
tasks easier and hence models as the one proposed here will gain
wide-spread applicability to many of the RNA-Seq eQTL studies to come.
%
%

\bibliographystyle{splncs03}
\bibliography{bibfile}

\end{document}